\documentclass[10pt,letterpaper]{article}
\usepackage{opex3}
\usepackage{amsmath}
\usepackage{amsfonts}

\begin{document}




\title{The multiparty coherent channel and its implementation with linear optics}

\author{Guangqiang He, Taizhi Liu and Xin Tao}

\address{State Key Lab of Advanced Optical Communication Systems and
Networks \\ Department of Electronic Engineering, Shanghai Jiaotong University, Shanghai
200240,China}
\email{gqhe@sjtu.edu.cn, taizhiliu88@gmail.com} 



\begin{abstract}
The continuous-variable coherent (conat) channel is a useful resource for coherent
communication, producing coherent teleportation and coherent
superdense coding. We extend the conat channel to multiparty conditions
by proposing definitions about multiparty
position-quadrature conat channel and multiparty
momentum-quadrature conat channel. We additionally provide two methods to
implement this channel using linear optics. One method is the
multiparty version of coherent communication assisted by
entanglement and classical communication (CCAECC). The other is
multiparty coherent superdense coding.
\end{abstract}



\section{Introduction}

The coherent bit (cobit) with discrete variables (DV) proposed by
Aram Harrow \cite{A. Harrow2004} is a powerful resource intermediate
between a quantum bit (qubit) and a classical bit (cbit). In \cite{A. Harrow2004},
a qubit channel can be described as: $|x\rangle^{A}\rightarrow|x\rangle^{B}$ ($\{|x\rangle\}_{x\in\{0,1\}}$ is a basis for $\mathbb{C}^2$. A is sender Alice. B is receiver Bob).
A cbit channel is: $|x\rangle^{A}\rightarrow|x\rangle^{B}|x\rangle^{E}$ (E is inaccessible environment).

A bipartite cobit
channel with DV can be described by the isometry:
$|x\rangle^{A}\rightarrow|x\rangle^{A}|x\rangle^{B}$.
For example, if Alice possesses an arbitrary qubit:
$|\psi\rangle^{A}=\alpha|0\rangle^{A}+\beta|1\rangle^{A}$ and transmits it through a
cobit channel, the channel generates: $|\phi\rangle^{AB}=\alpha|0\rangle^{A}|0\rangle^{B}+\beta|1\rangle^{A}|1\rangle^{B}$.
The process maintains the coherent superposition property of Alice's original state,
this is the reason for the channel's name\cite{A. Harrow2004}. In
\cite{M. M. Wilde2008}, Wilde and Brun compared and pointed out the
differences and connections among concepts about classical communication, quantum
communication, entanglement and cobit channel.

Recently, Wilde, Krovi, and Brun extended cobit channel with discrete
variables to continuous variables (CV), and they introduced the concept of
`conat channel'\cite{M. M. Wilde2007} as the CV counterpart of
the DV 'cobit channel'. In their work, Wilde and Brun proposed definitions of
ideal position-quadrature (PQ) and momentum-qudrature (MQ) conat channel
in Schroedinger-picture:
$|x\rangle^{A}\rightarrow|x\rangle^{A}|x\rangle^{B}$,$|p\rangle^{A}\rightarrow|p\rangle^{A}|p\rangle^{B}$
where $|x\rangle$ and $|p\rangle$ represents position and momentum eigenstate respectively.
Nonideal (finitely squeezed) conat channels are also discussed in the Heisenberg picture\cite{D. F. Walls}.

The coherent channel provides for coherent communication. Coherent
communication has several useful characteristics and applications.
It provides a coherent version of continuous-variable
teleportation and continuous-variable superdense coding, and they are proved dual under resource reversal \cite{A. Harrow2004,I. Devetak2006}.
In addition, it can be applied in remote state preparation (RSP),
consuming less entanglement than standard ways \cite{A. Harrow2004}.
Moreover, coherent communication also proves useful in error correcting
codes \cite{Brun2006,Brun2006arXiv}.

In this paper, we extend the notion of conat channel to multiparty situations
in order to perform multiparty coherent communication. Additionally, we
propose two methods of implementation with linear optical techniques and
analyze the noise accumulation in different EPR \cite{A. Einstein1935} resources.
Multiparty conat channel inherits all the useful
properties from bipartite conat channel.

Our paper is organized as follows. In Section 2, we provide a
general definition of multiparty position-quadrature (PQ) and momentum-qudrature (MQ) conat channel in
Heisenberg representation. In Section 3, two implementations of the
protocol with linear optical devices are outlined. A
brief conclusion is given in Section 4.

\section{Definitions of multiparty conat channel}

We propose a general definition of multiparty position-quadrature
(PQ) and momentum-qudrature (MQ) conat channel in Heisenberg representation\cite{D. F. Walls}.

The channel has only one sender: Alice and $n$ receivers: Alice, Bob, Charlie $\cdots$ and
Nick. Notice that the sender is also among the receivers. We denote sender Alice as $A$ and the received massage
after the channel as $A', B', C', \cdots, N'$.
The multiparty PQ conat channel $\tilde{\Delta}_{X}$ copies
the position quadrature of the sender to all the receivers with respective noise.
The resulting multimode state is
similar to multiparty Greenberger-Horne-Zeilinger (GHZ) entangled states\cite{D. M. Greenberger}.
The difference is that the total momentum of the
output modes is close to the original momentum of the sender
$\hat{p}_{A}$, encoding the transmitted message into all the
receivers involved in this channel, while the total momentum of GHZ
entangled state is zero. Multiparty MQ conat channel is similarly defined.

Definition 1: A multiparty PQ conat channel $\tilde{\Delta}_{X}$.

Mapping:
\begin{equation}\label{eq1}
[\hat{x}_{A}\ \hat{p}_{A}]^{T}\rightarrow[\hat{x}_{A'}\
\hat{p}_{A'}\ \hat{x}_{B'}\ \hat{p}_{B'}\ \cdots\ \hat{x}_{N'}\
\hat{p}_{N'}]^{T}
\end{equation}

Constraints:
\begin{equation}\label{eq2}
\begin{cases}
{\hat{x}_{A'}}=\hat{x}_{A}\\
\hat{x}_{B'}=\hat{x}_{A}+\hat{x}^{\ 1}_{\Delta_{X}}\\
\hat{x}_{C'}=\hat{x}_{A}+\hat{x}^{\ 2}_{\Delta_{X}}\\
\vdots\\
\hat{x}_{N'}=\hat{x}_{A}+\hat{x}^{n-1}_{\Delta_{X}}\\
\hat{p}_{A'}=\hat{p}_{A}+\hat{p}_{\Delta_{X}}
\end{cases}
\end{equation}
\begin{equation}\label{eq3}
\begin{cases}
\langle\hat{x}^{\ 1}_{\Delta_{X}}\rangle=\langle\hat{x}^{\ 2}_{\Delta_{X}}\rangle=\cdots=\langle\hat{x}^{n-1}_{\Delta_{X}}\rangle=0\\
\langle\hat{p}_{A'}+\hat{p}_{B'}+\hat{p}_{C'}+\cdots+\hat{p}_{N'}\rangle=\langle\hat{p}_A\rangle\Leftrightarrow
\langle\hat{p}_{\Delta_{X}}+\hat{p}_{B'}+\hat{p}_{C'}+\cdots+\hat{p}_{N'}\rangle=0\\
\langle(\hat{x}^{\ 1}_{\Delta_{X}})^{2}\rangle\leq\epsilon_{1}\\
\langle(\hat{x}^{\ 2}_{\Delta_{X}})^{2}\rangle\leq\epsilon_{2}\\
\vdots\\
\langle(\hat{x}^{n-1}_{\Delta_{X}})^{2}\rangle\leq\epsilon_{n-1}\\
\langle(\hat{p}_{\Delta_{X}}+\hat{p}_{B'}+\hat{p}_{C'}+\cdots+\hat{p}_{N'})^{2}\rangle\leq\epsilon_{n}
\end{cases}
\end{equation}
The canonical commutation relations of the Heisenberg-picture
observables are as follows:
\begin{equation}\label{eq4}
[\hat{x}_{A'},\hat{p}_{A'}]=[\hat{x}_{B'},\hat{p}_{B'}]=\cdots=[\hat{x}_{N'},\hat{p}_{N'}]=i
\end{equation}
Alice is the sender with mode \emph{A}, and $n$ receivers, from Alice to Nick, possess modes $A'$
through $N'$ respectively. Alice is the sender, as well as one of
the receivers. The constraints indicate that: Alice's position quadrature
remains unchanged, and it is copied to all the other receivers with the additional noise
$\hat{x}^{1}_{\Delta_{X}}$, $\hat{x}^{2}_{\Delta_{X}}$   $\cdots$
 $\hat{x}^{n-1}_{\Delta_{X}}$ respectively; the total momentum of the multimode state
is close to Alice's original momentum with a noise $\hat{p}_{\Delta_{X}}$.
The parameters $\epsilon_{1}$, $\epsilon_{2}$ $\cdots$ and $\epsilon_{n}$ in (3) determine the
performance of channel by bounding the noises presented.

Definition 2: A multiparty MQ conat channel $\tilde{\Delta}_{P}$.

Mapping:
\begin{equation}\label{eq5}
[\hat{x}_{A}\ \hat{p}_{A}]^{T}\rightarrow[\hat{x}_{A'}\
\hat{p}_{A'}\ \hat{x}_{B'}\ \hat{p}_{B'}\ \cdots\ \hat{x}_{N'}\
\hat{p}_{N'}]^{T}
\end{equation}

Constraints:
\begin{equation}\label{eq6}
\begin{cases}
{\hat{p}_{A'}}=\hat{p}_{A}\\
\hat{p}_{B'}=\hat{p}_{A}+\hat{p}^{\ 1}_{\Delta_{P}}\\
\hat{p}_{C'}=\hat{p}_{A}+\hat{p}^{\ 2}_{\Delta_{P}}\\
\vdots\\
\hat{p}_{N'}=\hat{p}_{A}+\hat{p}^{n-1}_{\Delta_{P}}\\
\hat{x}_{A'}=\hat{x}_{A}+\hat{x}_{\Delta_{P}}
\end{cases}
\end{equation}
\begin{equation}\label{eq7}
\begin{cases}
\langle\hat{p}^{\ 1}_{\Delta_{P}}\rangle=\langle\hat{p}^{\ 2}_{\Delta_{P}}\rangle=\cdots=\langle\hat{p}^{n-1}_{\Delta_{P}}\rangle=0\\
\langle\hat{x}_{A'}+\hat{x}_{B'}+\hat{x}_{C'}+\cdots+\hat{x}_{N'}\rangle=\langle\hat{x}_{A}\rangle\Leftrightarrow
\langle\hat{x}_{\Delta_{P}}+\hat{x}_{B'}+\hat{x}_{C'}+\cdots+\hat{x}_{N'}\rangle=0\\
\langle(\hat{p}^{\ 1}_{\Delta_{P}})^{2}\rangle\leq\epsilon_{1}\\
\langle(\hat{p}^{\ 2}_{\Delta_{P}})^{2}\rangle\leq\epsilon_{2}\\
\vdots\\
\langle(\hat{p}^{n-1}_{\Delta_{P}})^{2}\rangle\leq\epsilon_{n-1}\\
\langle(\hat{x}_{\Delta_{P}}+\hat{x}_{B'}+\hat{x}_{C'}+\cdots+\hat{x}_{N'})^{2}\rangle\leq\epsilon_{n}
\end{cases}
\end{equation}
The canonical commutation relations of the Heisenberg-picture observables are as follows:
\begin{equation}\label{eq8}
[\hat{x}_{A'},\hat{p}_{A'}]=[\hat{x}_{B'},\hat{p}_{B'}]=\cdots=[\hat{x}_{N'},\hat{p}_{N'}]=i
\end{equation}

\section{Implementations of multiparty conat channel using linear optics}

In this section, we outline two methods to implement the multiparty conat channel.
The first is multiparty coherent communication assisted by entanglement and classical communication (CCAECC) \cite{M. M. Wilde2008}.
The second is multiparty superdense coding which implements two multiparty coherent channels.

\textbf{Method 1:}

This method requires $(n+1)$-party GHZ
entangled state and classical communication channel. For simplicity, we implement a three-party PQ
conat channel as an demonstration. We will generalize to $n$-party PQ conat channel later
(multiparty MQ conat channel is similar).

In this method, Alice possesses a mode $A$ that she wants to transmit. Alice, Bob and Claire share a four-mode entangled state $A_1$, $A_2$, $B$ and $C$. $A_1$ and $A_2$ belong to Alice, $B$ belongs to Bob and $C$ belongs to Claire.

We use P. van Loock and S. L. Braunstein's protocol \cite{P. van
Loock2000} to generate a four-mode GHZ entanglement. This protocol starts from 4
original vacuum states ($\hat{x}_{1}^{(0)}$, $\hat{p}_{1}^{(0)}$, $\hat{x}_{2}^{(0)}$,
$\hat{p}_{2}^{(0)}$, $\hat{x}_{3}^{(0)}$, $\hat{p}_{3}^{(0)}$,
$\hat{x}_{4}^{(0)}$ and $\hat{p}_{4}^{(0)}$), squeezes them with
squeezing coefficients: $r_{1}$, $r_{2}$, $r_{3}$ and $r_{4}$
and generates entangled states $A_1$, $A_2$, $B$ and $C$.
The equations with coefficients calculated are given:
\begin{equation}\label{eq9}
\begin{cases}
\hat{x}_{A_1}&=\frac{1}{\sqrt{4}}e^{+r_{1}}\hat{x}_{1}^{(0)}+\sqrt{\frac{3}{4}}e^{-r_{2}}\hat{x}_{2}^{(0)}\\
\hat{p}_{A_1}&=\frac{1}{\sqrt{4}}e^{-r_{1}}\hat{p}_{1}^{(0)}+\sqrt{\frac{3}{4}}e^{+r_{2}}\hat{p}_{2}^{(0)}\\
\hat{x}_{A_2}&=\frac{1}{\sqrt{4}}e^{+r_{1}}\hat{x}_{1}^{(0)}-\frac{1}{\sqrt{12}}e^{-r_{2}}\hat{x}_{2}^{(0)}+\sqrt{\frac{2}{3}}e^{-r_{3}}\hat{x}_{3}^{(0)}\\
\hat{p}_{A_2}&=\frac{1}{\sqrt{4}}e^{-r_{1}}\hat{p}_{1}^{(0)}-\frac{1}{\sqrt{12}}e^{+r_{2}}\hat{p}_{2}^{(0)}+\sqrt{\frac{2}{3}}e^{+r_{3}}\hat{p}_{3}^{(0)}\\
\hat{x}_{B}&=\frac{1}{\sqrt{4}}e^{+r_{1}}\hat{x}_{1}^{(0)}-\frac{1}{\sqrt{12}}e^{-r_{2}}\hat{x}_{2}^{(0)}-\frac{1}{\sqrt{6}}e^{-r_{3}}\hat{x}_{3}^{(0)}+\frac{1}{\sqrt{2}}e^{-r_{4}}\hat{x}_{4}^{(0)}\\
\hat{p}_{B}&=\frac{1}{\sqrt{4}}e^{-r_{1}}\hat{p}_{1}^{(0)}-\frac{1}{\sqrt{12}}e^{+r_{2}}\hat{p}_{2}^{(0)}-\frac{1}{\sqrt{6}}e^{+r_{3}}\hat{p}_{3}^{(0)}+\frac{1}{\sqrt{2}}e^{+r_{4}}\hat{p}_{4}^{(0)}\\
\hat{x}_{C}&=\frac{1}{\sqrt{4}}e^{+r_{1}}\hat{x}_{1}^{(0)}-\frac{1}{\sqrt{12}}e^{-r_{2}}\hat{x}_{2}^{(0)}-\frac{1}{\sqrt{6}}e^{-r_{3}}\hat{x}_{3}^{(0)}-\frac{1}{\sqrt{2}}e^{-r_{4}}\hat{x}_{4}^{(0)}\\
\hat{p}_{C}&=\frac{1}{\sqrt{4}}e^{-r_{1}}\hat{p}_{1}^{(0)}-\frac{1}{\sqrt{12}}e^{+r_{2}}\hat{p}_{2}^{(0)}-\frac{1}{\sqrt{6}}e^{+r_{3}}\hat{p}_{3}^{(0)}-\frac{1}{\sqrt{2}}e^{+r_{4}}\hat{p}_{4}^{(0)}
\end{cases}
\end{equation}

And we assume that all squeezing coefficients equal to $r$.

\begin{figure}[htp1]
\begin{center}
\includegraphics [width=120mm]{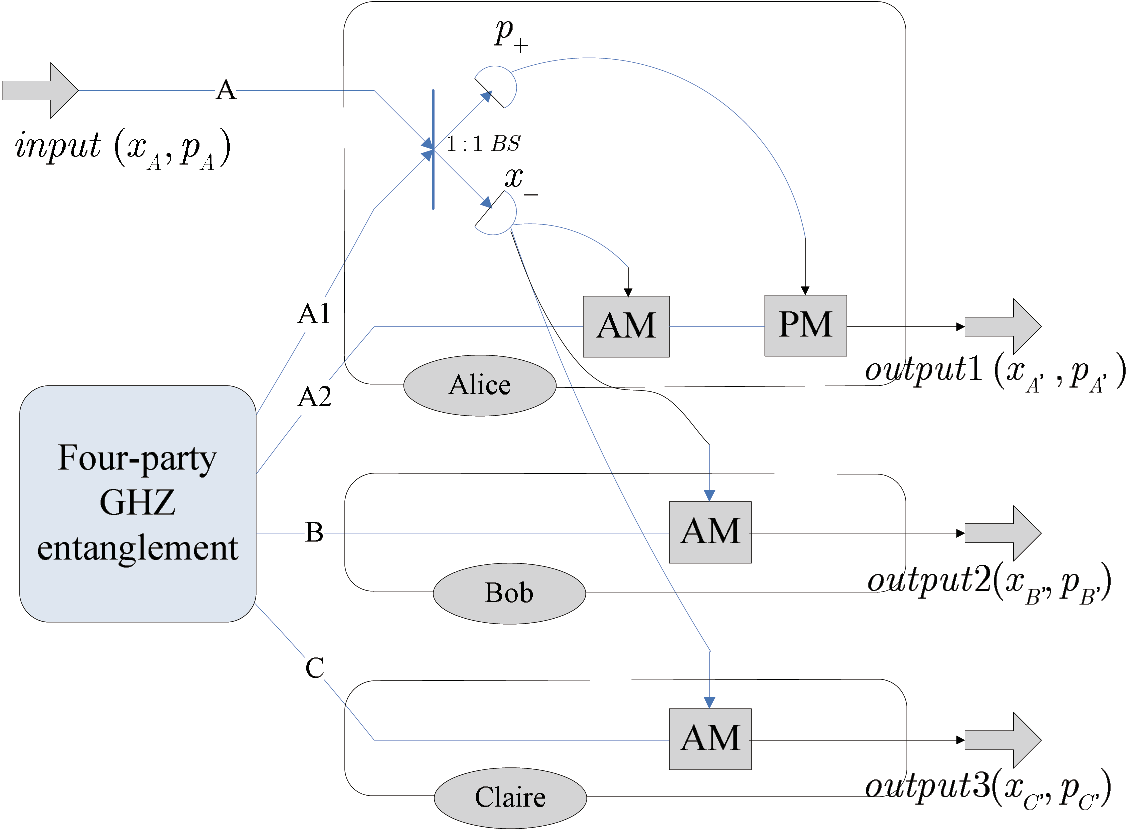}
\end{center}
\caption{ The three-party position-quadrature (PQ) protocol requires
a four-mode GHZ entangled state, and we label the four modes as
$A_{1}$, $A_{2}$, $B$, $C$. Alice possesses mode $A_{1}$ and $A_{2}$,
Bob possesses mode $B$ and Charlie possesses mode $C$.
Alice has an input mode $A$ with the message to be transmitted. The
elements used in this figure: \emph{AM} is an amplitude
modulator which displaces the position quadrature of an optical mode
\cite{M. M. Wilde2008}. \emph{PM} is a phase modulator which kicks
the momentum quadrature of an optical mode \cite{M. M. Wilde2008}. \emph{BS} is
a beam splitter.}\label{fig1}
\end{figure}

Then we present the transformations of the
operators in Heisenberg picture as follows:

\emph{Step 1. }Alice mixes mode $A$ and $A_{1}$ locally on a
balanced (50\%) beam splitter (BS), generating modes (+) and (-).
\begin{equation}\label{eq10}
\begin{cases}
\hat{x}_{+}=(\hat{x}_{A}+\hat{x}_{A_{1}})/\sqrt{2}\ ,\ \hat{p}_{+}=(\hat{p}_{A}+\hat{p}_{A_{1}})/\sqrt{2}\\
\hat{x}_{-}=(\hat{x}_{A}-\hat{x}_{A_{1}})/\sqrt{2}\ ,\ \hat{p}_{-}=(\hat{p}_{A}-\hat{p}_{A_{1}})/\sqrt{2}\\
\end{cases}
\end{equation}

\emph{Step 2. }\ \ We denote $\hat{x}_{A_{2}}$, $\hat{p}_{A_{2}}$,
$\hat{x}_{B}$ and $\hat{x}_{C}$ in terms of $\hat{x}_{-}$ and
$\hat{p}_{+}$.
\begin{equation}\label{eq11}
\begin{cases}
{\hat{x}_{A_{2}}}&=\hat{x}_{A}-\hat{x}_{A_{1}}+\hat{x}_{A_{2}}-\sqrt{2}\hat{x}_{-}\\
\hat{p}_{A_{2}}&=\hat{p}_{A}+(\hat{p}_{A_{1}}+\hat{p}_{A_{2}}+\hat{p}_{B}+\hat{p}_{C})-\hat{p}_{B}-\hat{p}_{C}-\sqrt{2}\hat{p}_{+}\\
\hat{x}_{B}&=\hat{x}_{A}-\hat{x}_{A_{1}}+\hat{x}_{B}-\sqrt{2}\hat{x}_{-}\\
\hat{x}_{C}&=\hat{x}_{A}-\hat{x}_{A_{1}}+\hat{x}_{C}-\sqrt{2}\hat{x}_{-}
\end{cases}
\end{equation}
Then Alice measures $\hat{x}_{-}$ and $\hat{p}_{+}$ through the
homodyne detection. After the homodyne detection, operator
$\hat{x}_{-}$ and $\hat{p}_{+}$ collapse to value $x_{-}$ and
$p_{+}$, then she sends the measurement value $x_{-}$ to Bob and Charlie over a classical communication channel. Suppose the photodetectors have efficiency $\eta$.

\emph{Step 3. }\ \ Alice displaces the position quadrature of her
mode $A_{2}$ by $\sqrt{2}x_{-}$ and her momentum quadrature by the
value of $\sqrt{2}p_{+}$, Bob and Charlie displace the position
quadrature of their possessed mode \emph{B} and \emph{C} respectively by
$\sqrt{2}x_{-}$, resulting mode $A'$, $B'$ and $C'$. After the modulations, we get:
\begin{equation}\label{eq12}
\begin{cases}
\hat{x}_{A'}&=\hat{x}_{A}-(\hat{x}_{A_{1}}-\hat{x}_{A_{2}})-\sqrt{2(1-\eta)/\eta}\hat{x}_{1}^{(0)}\\
\hat{p}_{A'}&=\hat{p}_{A}+(\hat{p}_{A_{1}}+\hat{p}_{A_{2}}+\hat{p}_{B}+\hat{p}_{C})-\hat{p}_{B}-\hat{p}_{C}+\sqrt{2(1-\eta)/\eta}\hat{p}_{2}^{(0)}\\
\hat{x}_{B'}&=\hat{x}_{A}-(\hat{x}_{A_{1}}-\hat{x}_{B})-\sqrt{2(1-\eta)/\eta}\hat{x}_{1}^{(0)}\\
\hat{p}_{B'}&=\hat{p}_{B}\\
\hat{x}_{C'}&=\hat{x}_{A}-(\hat{x}_{A_{1}}-\hat{x}_{C})-\sqrt{2(1-\eta)/\eta}\hat{x}_{1}^{(0)}\\
\hat{p}_{C'}&=\hat{p}_{C}
\end{cases}
\end{equation}
Finally, we can find the results we obtained satisfy the constraints
in (2) and (3), we get that
\begin{equation}\label{eq13}
\begin{cases}
\hat{x}^{\ 1}_{\Delta_{X}}=\hat{x}_{B'}-\hat{x}_{A}=\hat{x}_{B'}-\hat{x}_{A'}=\hat{x}_B-\hat{x}_{A_2}\\
\hat{x}^{\ 2}_{\Delta_{X}}=\hat{x}_{C'}-\hat{x}_{A}=\hat{x}_{C'}-\hat{x}_{A'}=\hat{x}_C-\hat{x}_{A_2}\\
\hat{p}_{\Delta_{X}}+\hat{p}_{B'}+\hat{p}_{C'}=(\hat{p}_{A_{1}}+\hat{p}_{A_{2}}+\hat{p}_{B}+\hat{p}_{C})+\sqrt{2(1-\eta)/\eta}\hat{p}_{2}^{(0)}\\
\langle\hat{x}^{\ 1}_{\Delta_{X}}\rangle=\langle\hat{x}^{\ 2}_{\Delta_{X}}\rangle=\langle\hat{p}_{\Delta_{X}}+\hat{p}_{B'}+\hat{p}_{C'}\rangle=0\\
\langle(\hat{x}^{\ 1}_{\Delta_{X}})^2\rangle\leq2e^{-2r},\ \langle(\hat{x}^{\ 2}_{\Delta_{X}})^2\rangle\leq2e^{-2r},\ \\
\langle(\hat{p}_{\Delta_{X}}+\hat{p}_{B'}+\hat{p}_{C'})^2\rangle\leq4e^{-2r}+2(1-\eta)/\eta
\end{cases}
\end{equation}
So, parameters $\epsilon_{1}$, $\epsilon_{2}$ and $\epsilon_{3}$ in constraints (3) are:
\begin{equation}
\epsilon_{1}=\epsilon_{2}=2e^{-2r},\
\epsilon_{3}=4e^{-2r}+2(1-\eta)/\eta
\end{equation}
In this case , \textit{n} is equal to 3. As \textit{n} increases,
$\epsilon_{1}$, $\epsilon_{2}$ $\cdots$ $\epsilon_{n-1}$ remain
unchanged and $\epsilon_{n}$ amounts to
$(n+1)e^{-2r}+2(1-\eta)/\eta$. When it comes to the general
\emph{n}-party PQ conat channel in this method, we can implement it
easily and similarly using the $(n+1)$-party GHZ entanglement.

The implementation of multiparty MQ conat channel is similar to
multiparty PQ conat channel, viewing the operator $\hat{x}$ as
$\hat{p}$ and operator $\hat{p}$ as $\hat{x}$. The required
resource is a GHZ-like entanglement state with total position
$\hat{x}_{A_{1}}+\hat{x}_{A_{2}}+\hat{x}_{B}+\cdots+\hat{x}_{N}\rightarrow0$
and relative momenta equal. So we omit these discussions here.

\textbf{Method 2:}

Widle, Krovi and Brun proposed a protocol of coherent superdense
coding recently \cite{M. M. Wilde2007}. Inspired by their work, we
provide a multiparty version of coherent superdense coding. The
protocol is equivalent to multiparty conat channels, a
multiparty PQ conat channel and a multiparty MQ conat channel. In
this method, the channel has only one sender which has two
modes to be transmitted, and finally has $n$ receivers obtained the
two modes respectively. In addition, $(n-1)$ prepared EPR pairs among the receivers are
required.

We also use P. van Loock and S. L. Braunstein's method to generate an
EPR pair. In Heisenberg representation:
\begin{equation}\label{eq14}
\begin{cases}
\hat{x}_{1}=(e^{+r}\hat{x}_{1}^{(0)}+e^{-r}\hat{x}_{2}^{(0)})/\sqrt{2},\
&\hat{p}_{1}=(e^{-r}\hat{p}_{1}^{(0)}+e^{+r}\hat{p}_{2}^{(0)})/\sqrt{2}\\
\hat{x}_{2}=(e^{+r}\hat{x}_{1}^{(0)}-e^{-r}\hat{x}_{2}^{(0)})/\sqrt{2},\
&\hat{p}_{2}=(e^{-r}\hat{p}_{1}^{(0)}-e^{+r}\hat{p}_{2}^{(0)})/\sqrt{2}
\end{cases}
\end{equation}

Local quantum nondemolition (QND) interactions are also employed. After the
transformations, we get\cite{R. Filip2005}:
\begin{equation}\label{eq15}
\begin{cases}
\hat{x}_{1'}=\hat{x}_{1},\
&\hat{p}_{1'}=\hat{p}_{1}-\hat{p}_{2}\\
\hat{x}_{2'}=\hat{x}_{1}+\hat{x}_{2},\ &\hat{p}_{2'}=\hat{p}_{2}
\end{cases}
\end{equation}

The QND interaction with a phase adjust can be described as\cite{R. Filip2005}:
\begin{equation}\label{eq16}
\begin{cases}
\hat{x}_{1'}=\hat{x}_{1}-\hat{x}_{2},\
&\hat{p}_{1'}=\hat{p}_{1}\\
\hat{x}_{2'}=\hat{x}_{2},\ &\hat{p}_{2'}=\hat{p}_{1}+\hat{p}_{2}
\end{cases}
\end{equation}

\begin{figure}[htp1]
\begin{center}
\includegraphics [width=90mm]{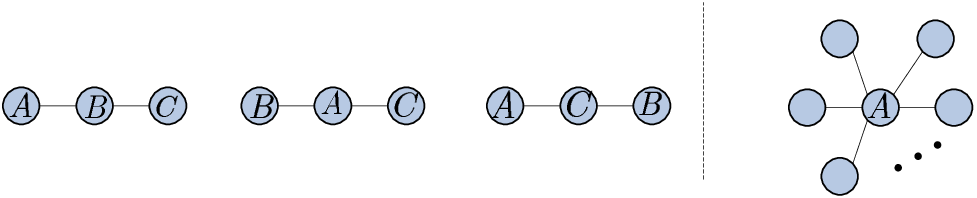}
\end{center}
\caption{ On the left side of the vertical line, we present all the
scenarios of graph which represents the prepared entanglement
resources when n is equal to 3. The number of total scenarios
is three. On the right side of the vertical line, the graph
indicates the best prepared entanglement resources for n-party conat
channel.} \label{fig2}
\end{figure}

Our protocol requires $(n-1)$ EPR pairs for receivers. We
illustrate these requirements in Figure \ref{fig2}, on the condition that $n$ equals to 3. We use a graph to
illustrate the entanglement relations among the parties involved. A vertex in the graph represents an individual party in
he channel, and the edge between two vertices indicates EPR entanglement relation
between two parties. We will introduce concepts from graph theory: when two vertices are the terminals
of a edge, they are called 'adjacent'; we refer two
vertices as connected when a path exists between them; a connected
graph is a graph in which any two of the vertices are connected.
This method requires that the graph of
entanglement resources be a connected graph. The channel has $n$ parties involved. The $(n-1)$
EPR pairs prepared among the $n$ parties ensure that the
representing graph is connected, and any two vertices of the graph
 has only one path.

Without loss of generalization, we discuss two scenarios
in Figure \ref{fig2}. Scenario 1 corresponds to the first graph in
Figure \ref{fig2}, Scenario 2 corresponds to the second graph.
Since $n$ is equal to 3, the channel involves one sender and three
receivers, two EPR pairs is required.

\begin{figure}[htp1]
\begin{center}
\includegraphics [width=120mm]{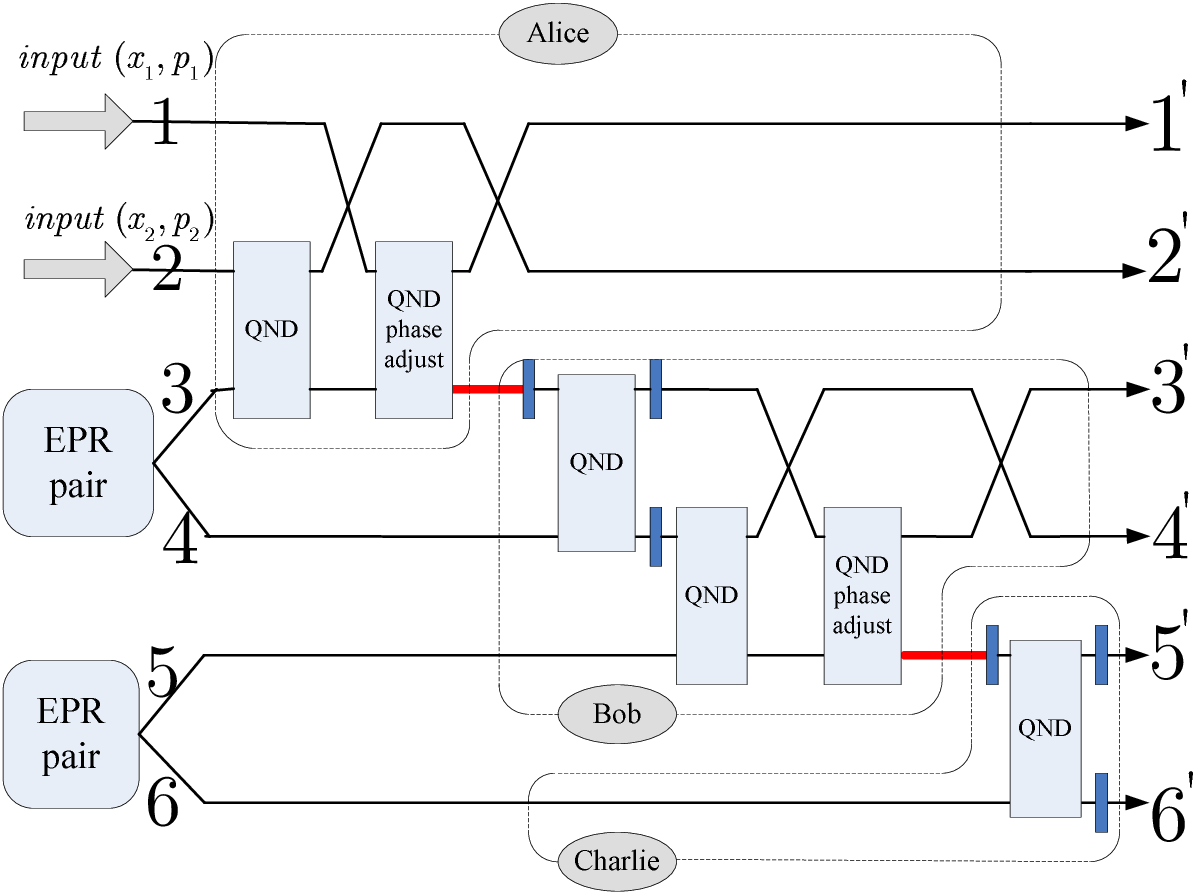}
\end{center}
\caption{This figure outlines our scheme. The thick red line in this
figure represents a quantum channel between two parties. The
local operations and modes are enclosed by the dashed with the name
of the party. The The blue thin rectangle means the phase shifter
about $\pi$.} \label{fig3}
\end{figure}

\emph{Scenario 1. } In this scenario, Alice and Bob shares an EPR
entanglement pair: mode 3 for Alice and mode 4 for
Bob, Bob and Charlie share an EPR entanglement pair: mode 5 for Bob and mode 6 for Charlie.
Alice possesses mode 1 and 2 which are to be transmitted. Figure \ref{fig3} gives the schematic linear optics
circuit.

In step 1, Alice couples her mode 2 and 3 in QND interaction, and
then couples mode 1 and 3 in QND phase adjust interaction. In step
2, Alice sends her mode 3 to Bob through quantum channel, so Bob
now possesses three modes, mode 3, 4 and 5. Then he performs a
series of QND interactions: first couples mode 3, 4, then mode 4 and 5, finally QND phase adjust interaction about
mode 3, 5. In step 3, Bob sends mode 5 to Charlie through a
quantum channel. Charlie couples his two modes.
then we can get the resulting modes in
Heisenberg picture.

\begin{equation}\label{eq17}
\begin{cases}
\hat{x}_{1'}=\hat{x}_{1}-(\hat{x}_{2}+\hat{x}_{3})\ , \ &\hat{p}_{1'}=\hat{p}_{1}\\
\hat{x}_{2'}=\hat{x}_{2}\ ,\ &\hat{p}_{2'}=\hat{p}_{2}-\hat{p}_{3}\\
\hat{x}_{3'}=\hat{x}_{2}+\hat{x}_{3}-(\hat{x}_{2}+\hat{x}_{3}-\hat{x}_{4}+\hat{x}_{5})\,\ &\hat{p}_{3'}=\hat{p}_{1}+\hat{p}_{3}+\hat{p}_{4}\\
\hat{x}_{4'}=\hat{x}_{2}+\hat{x}_{3}-\hat{x}_{4}\ ,\
&\hat{p}_{4'}=-\hat{p}_{4}-\hat{p}_{5}\\
\hat{x}_{5'}=\hat{x}_{2}+\hat{x}_{3}-\hat{x}_{4}+\hat{x}_{5}\ , \
&\hat{p}_{5'}=\hat{p}_{1}+\hat{p}_{3}+\hat{p}_{4}+\hat{p}_{5}+\hat{p}_{6}\\
\hat{x}_{6'}=\hat{x}_{2}+\hat{x}_{3}-\hat{x}_{4}+\hat{x}_{5}-\hat{x}_{6}\
, \ &\hat{p}_{6'}=-\hat{p}_{6}
\end{cases}
\end{equation}
Modes 1', 3', 5' implement a three-party MQ conat channel by
satisfying the constraints in definition 2; and modes 2', 4', 6'
satisfy definition 1 and work as a three-party PQ conat channel.

\begin{figure}[htp1]
\begin{center}
\includegraphics [width=120mm]{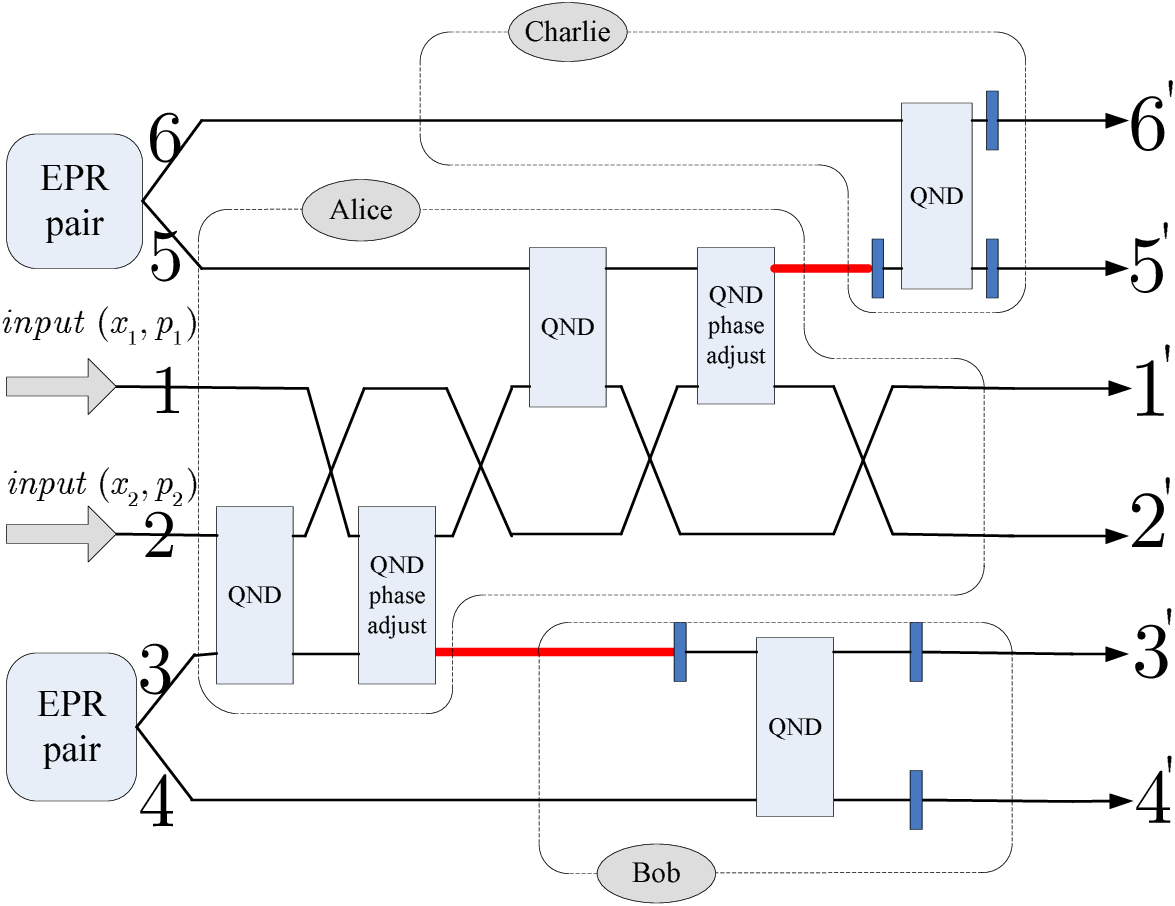}
\end{center}
\caption{The thick red line in this figure represents a quantum
channel between two parties. The local operations and modes are
enclosed by the dashed with the name of the party. The The blue thin
rectangle means the phase shifter about $\pi$.}
\label{fig4}
\end{figure}

\emph{Scenario 2. } In this scenario, Alice possesses four modes
at the beginning:
 mode 1, 2, 3, 5, Bob possesses mode 4 and
Charlie possesses mode 6. Mode 1 and 2 is to be transmitted while modes 3, 4, 5, 6 are the auxiliary modes.
Modes 3, 4 are EPR pair, as well as modes 5, 6. We give Figure
\ref{fig4} to describe this protocol.

In step 1, Alice performs a series of QND interactions: first, couples mode 2, 3; then QND phase adjust interaction about mode
1, 3; then QND interaction between mode 2, 5; finally QND phase adjust
interaction between mode 1, 5. In step 2, Alice sends mode 3 to Bob and
mode 5 to Charlie through quantum channels. Then Bob possesses two
modes 3, 4 and Charlie possesses modes 5, 6. They perform local QND
interactions on their two modes respectively.
The resulting modes are given as follows:

\begin{equation}\label{eq18}
\begin{cases}
\hat{x}_{1'}=\hat{x}_{1}-(\hat{x}_{2}+\hat{x}_{3})-(\hat{x}_{2}+\hat{x}_{5})\ , \ &\hat{p}_{1'}=\hat{p}_{1}\\
\hat{x}_{2'}=\hat{x}_{2}\ ,\
&\hat{p}_{2'}=\hat{p}_{2}-\hat{p}_{3}-\hat{p}_{5}\\
\hat{x}_{3'}=\hat{x}_{2}+\hat{x}_{3}\ ,\
&\hat{p}_{3'}=\hat{p}_{1}+\hat{p}_{3}+\hat{p}_{4}\\
\hat{x}_{4'}=\hat{x}_{2}+\hat{x}_{3}-\hat{x}_{4}\ ,\
&\hat{p}_{4'}=-\hat{p}_{4}\\
\hat{x}_{5'}=\hat{x}_{2}+\hat{x}_{5}\ , \
&\hat{p}_{5'}=\hat{p}_{1}+\hat{p}_{5}+\hat{p}_{6}\\
\hat{x}_{6'}=\hat{x}_{2}+\hat{x}_{5}-\hat{x}_{6}\ , \
&\hat{p}_{6'}=-\hat{p}_{6}
\end{cases}
\end{equation}

As in scenario 1, the output modes 1', 3', 5' perform a three-party
MQ conat channel and modes  2', 4', 6' perform a a three-party PQ
conat channel. So two multiparty conat channels are produced in this
protocol.

Then we discuss the noise of the channels in these two scenarios,
we assume that the local QND interaction is ideal. So we get that

\emph{In Scenario 1}:
\begin{equation}\label{eq19}
\begin{cases}
PQ\ conat\ channel&:\ \epsilon_{1}=2e^{-2r},\
\epsilon_{2}=4e^{-2r},\ \epsilon_{3}=4e^{-2r}\\
MQ\ conat\ channel&:\ \epsilon_{1}=2e^{-2r},\ \epsilon_{2}=4e^{-2r},\
\epsilon_{3}=0
\end{cases}
\end{equation}

\emph{In Scenario 2}:
\begin{equation}\label{eq20}
\begin{cases}
PQ\ conat\ channel&:\ \epsilon_{1}=2e^{-2r},\
\epsilon_{2}=2e^{-2r},\ \epsilon_{3}=4e^{-2r}\\
MQ\ conat\ channel&:\ \epsilon_{1}=2e^{-2r},\ \epsilon_{2}=2e^{-2r},\
\epsilon_{3}=0
\end{cases}
\end{equation}

Comparing these two scenarios, we find that the noise of
scenario 2 is lower. We can reach a conclusion that the longer
the path between one party and Alice, the larger the noise of the
party and of the channel. As you
see in Figure \ref{fig2}, the first graph represents Scenario 1,
the second graph represents Scenario 2. For instance, in
Scenario 1, the length of the path between Bob and Alice is one,
and the length of the path between Charlie and Alice is two. So
Charlie gets larger noise $\epsilon_{2}=4e^{-2r}$. The longer the
path between one party and Alice, the larger the accumulation of the
noise this party gets. In Scenario 2, the length of the path
between Alice and any party is one, there is no accumulation of the
noise, so Scenario 2 is better. For n-party conat channel, the best
prepared entanglement resources is the rightmost graph in Figure
\ref{fig2}.

\section{Conclusion}
Coherent bits (cobits) are intermediate in power between
qubits and cbits. Qubit sources can be used to simulate cobit sources, and cobit sources can simulate cbit sources\cite{A. Harrow2004}. Coherent communications offers a new view of quantum information elements.

In this paper, we extended the notion of continuous-variable
coherent (conat) channel to multiparty conditions and proposed two definitions of
it. Then we propose two implementations of multiparty conat channel
using linear optics. One method is the multiparty version of
coherent communication assisted by entanglement and classical
communication (CCAECC). The other is multiparty coherent superdense
coding which implements two multiparty coherent channels. We
also discuss the noise of the channel in two scenarios when n equals to 3.

\section{Acknowledgements}

This work was supported by the National Natural Science Foundation of China (Grants No. 61102053), the Scientific Research
Foundation for the Returned Overseas Chinese Scholars, State Education Ministry, SMC Excellent Young Faculty program (2011), SJTU Young Teacher Foundation (Grants No. A2831B) and SJTU PRP (Grants No. T03013002).

\end{document}